# Discovering the Growth Histories of Exoplanets: The Saturn Analog HD 149026b

Short title: The growth of HD 149026b

Sarah E. Dodson-Robinson[1]

*NASA Exoplanet Science Institute, California Institute of Technology*

*770 S. Wilson Ave, Pasadena, CA 91125*

`sdr@ipac.caltech.edu`

Peter Bodenheimer

*UCO/Lick Observatory, University of California at Santa Cruz*

*1156 High St., Santa Cruz, CA 95064*

---

[1] Formerly Sarah E. Robinson

# ABSTRACT

The transiting "hot Saturn" HD 149026b, which has the highest mean density of any confirmed planet in the Neptune-Jupiter mass range, has challenged theories of planet formation since its discovery in 2005. Previous investigations could not explain the origin of the planet's 45-110 Earth-mass solid core without invoking catastrophes such as gas giant collisions or heavy planetesimal bombardment launched by neighboring planets. Here we show that HD 149026b's large core can be successfully explained by the standard core accretion theory of planet formation. The keys to our reconstruction of HD 149026b are (1) applying a model of the solar nebula to describe the protoplanet nursery; (2) placing the planet initially on a long-period orbit at Saturn's heliocentric distance of 9.5 AU; and (3) adjusting the solid mass in the HD 149026 disk to twice that of the solar nebula in accordance with the star's heavy element enrichment. We show that the planet's migration into its current orbit at 0.042 AU is consistent with our formation model. Our study of HD 149026b demonstrates that it is possible to discover the growth history of any planet with a well-defined core mass that orbits a solar-type star.

*Key words:* planetary systems: formation – planetary systems: protoplanetary disks – planets and satellites: formation – methods: analytical

# 1. INTRODUCTION: CORE ACCRETION AND EXOPLANET FORMATION

The most widely accepted explanation of gas giant growth is the core accretion theory, in which icy planetesimals embedded in a protostellar disk coalesce to form a solid planet core (Safronov 1969; Pollack et al. 1996). Upon reaching ~15 $M_\oplus$ (Earth mass), the solid core gravitationally destabilizes the surrounding hydrogen and helium gas to accrete a massive atmosphere (Papaloizou & Nelson 2005). Planets forming in icy regions at ≥ 5 AU from their parent stars may then migrate into orbits at < 0.1 AU, becoming "hot Jupiters" (Lin et al. 1996). The theory's most serious problem—the fact that the time required to form Jupiter, Saturn, Uranus and Neptune initially appeared longer than typical protostellar disk lifetimes (Pollack et al. 1996)—has recently been resolved for Jupiter and Saturn (Alibert et al. 2005; Hubickyj et al. 2005; Dodson-Robinson et al. 2008).

However, most core accretion simulations predict core masses of ≤15 $M_\oplus$—far less than the 45-110 $M_\oplus$ required to match the observed mean density of the solid-rich exoplanet HD 149026b. One previous investigation of HD 149026b's growth proposed that the large core was due to either a gas-ejecting collision between two progenitor giant planets, or an unusually high planetesimal flux launched by perturbations from a companion planet (Ikoma et al. 2006). Another scenario combined a low-pressure solar nebula with an unusually large planetesimal flux of $10^{-2}$ $M_\oplus$ yr$^{-1}$—about a moon mass per

year—to allow the planet to form in its current orbit at 0.042 AU (Broeg & Wuchterl 2007). Although planetary systems are chaotic and catastrophic collisions cannot be ruled out, our goal is to bring as much determinism as possible to planet formation theory by calculating the solid/gas mass ratio, formation timescale and orbital evolution of exoplanets using the only the standard core accretion-migration paradigm.

A core accretion simulation calculates formation timescale, solid core mass, and gaseous atmosphere mass based on the planet's initial distance $R_0$ from its host star and the surface density $\Sigma$ of planetesimals available for core formation. Unfortunately for the predictive power of the theory, planets' resonant interactions with their natal protostellar disks trigger radial migration (Ward 1997; Nelson et al. 2000; Papaloizou & Nelson 2005) that makes direct measurement of $R_0$ impossible. However, most known exoplanets orbit solar-type (FGK) stars with similar masses and evolutionary histories to the sun. We can therefore use models of the solar nebula (the sun's protostellar disk), which contain theoretical predictions of $\Sigma(R)$, to recreate the initial conditions for planet formation around sunlike stars.

## 2. THE SATURN ANALOG HD 149026b

The discoverers of the exoplanet HD 149026b dubbed the planet a "hot Saturn" because of its mass, 1.2 Saturn masses, and the intense radiation it absorbs due to its position at only 0.042 AU from the G7 IV star HD 149026 (Sato et al. 2005). Optical photometry of the transit of HD 149026b led to a measured radius of 0.85 Saturn radii (Sato et al. 2005; Charbonneau et al. 2006; Winn et al. 2008). Nutzman et al. (2008)

measured a slightly larger transit depth in the *Spitzer* IRAC 8-μm band, leading to a radius estimate of 0.90 ± 0.048 Saturn radii, which is consistent with the optically measured radius within uncertainties. Recent NICMOS transit observations by Carter et al. (2009) point to an increased value of the star radius, 1.541 $R_\odot$, and a planet radius of 0.96 $R_{Saturn}$. The transit photometry and radial-velocity mass measurements therefore indicate that HD 149026b can be considered a dense version of Saturn. Despite the dissimilar orbits that Saturn and HD 149026b planets now occupy (Saturn is located 9.5 AU from the sun), migration allows for the possibility that HD 149026b formed far from its parent star, in an orbit similar to Saturn's.

Gravitational moment measurements place Saturn's core mass in the range 9-22 $M_\oplus$ (Saumon & Guillot 2004). Due to uncertainties in the planet radius and composition, the core mass of HD 149026b, though unequivocally large, is less well constrained. Based on the optically measured radius, Baraffe et al. (2008) derive a heavy element mass range of 60-80 $M_\oplus$ for core compositions ranging from pure rock to pure water. Using non-gray atmosphere models and high-pressure equations of state, Fortney et al. (2006) find that the HD 149026b's total heavy element abundance (core and enevelope) could be anywhere between 60 and 93 $M_\oplus$. Exploring the possibility of heavy element-enriched planet atmospheres with metal abundances up to 10x Solar, Burrows et al. (2007) derived a higher core mass of 80-110 $M_\oplus$. Finally, Carter et al. (2009) used the grids of planetary cooling and contraction models calculated by Sato et al. (2005) and Fortney et al. (2007) in combination with the larger NICMOS-measured planet radius to derive a core mass of 45-70 $M_\oplus$. Plausible core mass estimates of HD 149026b therefore range from 45-110 $M_\oplus$, making it at least twice as massive as Saturn's core.

There is one important distinction between the sun and HD 149026 that should account for the different mean densities of their respective "Saturns": HD 149026 contains an unusually high abundance of heavy elements. Spectral synthesis modeling gives [Fe/H] = 0.36, indicating that HD 149026 is more iron-rich than the sun by a factor of 2.3 (Sato et al. 2005). Such iron-rich stars are rare: of the 1040 solar-type stars observed in the Spectroscopic Properties of Cool Stars survey (Valenti & Fischer 2005), only 89 have more than twice the solar iron abundance. Statistical analyses of the exoplanet population have shown that the probability of planet detection increases with stellar iron, silicon and nickel abundance (Fischer et al. 2005, Robinson et al. 2006). Furthermore, the models of Guillot et al. (2006) and Burrows et al. (2007) suggest that planets with large amounts of heavy elements tend to be associated with metal-rich stars. We therefore expect any Saturn-like planet orbiting HD 149026 to contain a higher proportion of solids than Saturn itself.

### 3. INITIAL CONDITIONS FOR PLANET FORMATION

To reconstruct the growth of HD 149026b, we need reasonable guesses for $R_0$ and $\Sigma$. At 1.3 solar masses, HD 149026b is essentially a sunlike star, so we begin by assuming its protostellar disk had the same total mass (gas+solid) as the solar nebula. We reconstruct the solid surface density profile $\Sigma(R)$ of the HD 149026b disk by scaling the Σ(R) curves predicted for the solar nebula (Dodson-Robinson et al. 2009) upward by a factor of two. Fig. 1 shows time shapshots of Σ(R) during the planetesimal-building epoch of the HD 149026 disk. The required estimate of $R_0$ can be obtained by comparing the theoretical "isolation mass"—the total mass a planet core can reach if it accretes all the solids in its feeding zone—with HD 149026b's actual core mass. The isolation mass (Lissauer 1993) is given by

$$\frac{M_{iso}}{M_\oplus} = (2.1\times 10^{-3})\Sigma^{3/2}\left(\frac{R}{1AU}\right)^3 \quad (1)$$

where Σ is the solid surface density of planetesimals in g cm$^{-2}$ at the beginning of planet growth, R is the distance from the star, and $M_{iso}$ is the isolation mass. We see from Eq. 1 that the two ways to increase isolation mass are to raise Σ by constructing a solid-rich disk (which we have done, based on the supersolar metallicity of HD 149026), and—most importantly, due to the $R^3$ scaling—to place the planet on a long-period orbit. Note that the planet is not required to reach isolation mass—unless planet formation is a 100% efficient process with no solid body loss, the feeding zone will probably contain more planetesimals than necessary for building the core.

Using the above equation and the Σ(R) curves constructed for the HD 149026 disk, we find that planet's smallest possible feeding zone distance, if the core takes on the minimum value of 45 $M_\oplus$ and reaches isolation mass, is 6.0 AU. To match the higher core mass values derived from the optical photometry, HD 149026b had to begin forming a minimum of 8.0 AU from the star. The closest match to this orbit in the solar system is Saturn, located 9.5 AU from the sun. Our working hypothesis, based on the structural similarity between HD 149026b and Saturn and the possible locations of HD 149026b's feeding zone, is that the two planets share a formation pathway.

## 4. SIMULATION RESULTS AND DISCUSSION

We set up an experiment to test our hypothesis that the same model protostellar disk that produced Saturn, containing 8.6 g cm$^{-2}$ of planetesimals at 9.5 AU (Dodson-Robinson et al. 2008; 2009), with surface density doubled to match the heavy element content of HD 149026, could account for HD 149026b's large core. In a successful outcome, the planet would reach its current mass within a typical disk lifetime of 2-3 Myr (Haisch et al. 2001) and contain the observed proportion of solid core and atmospheric gas. We placed a 1 $M_\oplus$ solid planet embryo at Saturn's heliocentric distance, 9.5 AU, in the HD 149026 disk. A stellar evolution code adapted to the core accretion and gas capture process was coupled to the simulated solar nebula to track the planet's growth. The numerical methods used here are described extensively in the literature (Pollack et al. 1996; Laughlin et al. 2004; Dodson-Robinson et al. 2008; Dodson-Robinson et al. 2009). To account for the time it would take to form the 1 $M_\oplus$ seed embryo, we introduced a 0.15-Myr time lag between protostellar disk formation and the onset of planet growth (Dodson-Robinson et al. 2008). We ended the calculation once HD 149026b reached its current mass of 115 $M_\oplus$.

The solid surface density $\Sigma$ at (t = 0.15 Myr, $R_0$ = 9.5 AU) is 17.2 g cm$^{-2}$, which is a factor of 17 higher than the minimum-mass solar nebula (MMSN; Weidenschilling 1977). This deviation from the canonical MMSN reflects both the metal-rich nature of HD 149026 and the fundamentally inefficient nature of planet formation. Solid depletion processes include loss of meter-sized rocks to gas drag and collisional grinding down of planetesimal belts followed by radiation-driven small grain loss. Even planet cores in regions without active mass loss processes may never reach isolation mass or deplete their feeding zones. The "extra" planetesimals will likely be scattered into structures like the Oort cloud, which contains ~$10^{12}$ comets of radius >1 km for a total mass of >50 $M_\oplus$ (Lissauer 1993; Weissman 1996). Theoretical protostellar collapse calculations (e.g. Yorke & Bodenheimer 1999) predict that the initial solar nebula was at least 10x more massive than the MMSN. The MMSN, reflecting as it does the final outcome of planet growth, should not be used to set the initial conditions for core accretion.

Fig. 2 shows the solid mass, gas mass and total mass of HD 149026b as a function of time during the planet's growth epoch. The planet reaches its current mass in only 1.57 Myr, which is comfortably below the 2-3 Myr median disk lifetime. At t = 1.57 Myr, the planet's solid core contains 66 $M_\oplus$, which is in the middle of the range of possible core masses inferred from transit photometry and planetary evolution models (see §2). Clearly, our simulation passes both the formation time test and the interior structure test.

The onset of runaway gas accretion occurs at t ~ 1.5 Myr, when the solid core reaches ~45 $M_\oplus$. At first glance, this seems like an extraordinarily high critical core mass compared to the canonical value of 10 $M_\oplus$ derived by Mizuno (1980) and the 15 $M_\oplus$ quoted by Papaloizou & Nelson (2005). However, critical core mass is really a function of orbital radius and solid/gas ratio. Rafikov (2006) derived the functional form of the critical core mass-semimajor axis relationship for a disk with $\Sigma(R) \propto R^{-3/2}$. Near the star, where gas density is high, a core with mass <7 $M_\oplus$ can initiate Kelvin-Helmholz contraction of the surrounding gas. As gas density decreases with distance from the star, critical core mass increases as $M_{cr} \propto R^{5/8}$. This power law continues until a turnover radius at which the planetesimal accretion rate becomes too low to stabilize the protoplanetary envelope against collapse. Beyond the turnover radius, the critical core mass decreases as $M_{cr} \propto R^{-\alpha}$, where $9/10 \leq \alpha \leq 9/5$. The numerical simulations of Rafikov (2006) show that if planetesimal accretion rates are high, the critical core mass reaches 90 $M_\oplus$ at the turnover radius, ~15 AU. At 9.5 AU, the critical core mass varies from 20-80 $M_\oplus$ depending on planetesimal flux.

Ikoma et al. (2006) explored the possibility that even though the HD 149026b likely contains >60 $M_\oplus$ of solids, its core never reached critical mass. They note that the core could have been subcritical even up to 50-80 $M_\oplus$ if the planet formed on a wide orbit and accreted planetesimals at a high rate. However, they rejected the hypothesis that planet never reached runaway gas accretion based on the ratio of core mass to total mass:

planets with subcritical cores tend to have 70% or more of their mass in the core, and HD 149026b has a core mass fraction less than 70% in all but the most extreme planetary evolution models. By placing HD 149026b on a Saturn-type orbit in a solid-rich disk, we have constructed similar conditions to the Ikoma et al. subcritical core scenario. However, we allow the planet to pass critical core mass and accrete a substantial gaseous envelope. Furthermore, our core continues to accrete solids beyond the critical value simply because the critical core mass is less than the isolation mass of 128 $M_\oplus$.

We have yet to explain the planet's migration from $R_0$ = 9.5 AU to its present position at only 0.042 AU. Although migration is not explicitly included in our calculations, we can still make a heuristic estimate of whether our results allow for migration and when it occurs. We consider Type II migration (Nelson et al. 2000), which begins when the planet opens a gap in the disk gas surrounding its own orbit. Gap opening occurs once the planet's Hill sphere radius is equal to the disk's scale height at the planet location. Fig. 3 shows the concurrent evolution of the disk pressure scale height and the planet's Hill radius. Gap opening and the onset of Type II migration occur at t=1.45 Myr, which is 0.12 Myr before the planet is fully assembled. Literature estimates of the Type II migration timescale are ~0.1 Myr (Nelson et al. 2000; Papaloizou & Nelson 2005), in excellent agreement with our calculated Type II migration epoch. Once the planet migrates into the inner magnetospheric cavity of the protostellar disk, typically ~0.08 AU, the torques provided by gas exterior to the orbit vanish and the radial motion halts (Lin et al. 1996). Type II migration occurring from t=1.45-1.57 Myr could explain why HD 149026b reaches only 115 $M_\oplus$ even though 9 g cm$^{-2}$ of solid surface density remain in its original growth zone at 9.5 AU at the end of our simulation. Once the planet has migrated into the inner disk cavity, it no longer has access to that solid material and cannot continue to grow.

Although Broeg & Wuchterl (2007) and Bodenheimer et al. (2000) have shown that it is, in principle, possible to form Saturn-mass planets with large cores *in situ* at distances $R < 0.05$ AU, these models rely on several arbitrary and special assumptions. First, because the Hill radius is so small near the star, isolation masses of close-in planets are sub-moon-mass for any reasonable nebula density and the required 0.1 $M_\oplus$ seed embryo would have had to form at a larger distance and then migrate to $R < 0.05$ AU. Once the embryo stopped migrating—perhaps because it entered the magnetospheric cavity—it would be bombarded with a planetesimal flux between $10^{-5}$ $M_\oplus$ yr$^{-1}$ and $10^{-2}$ $M_\oplus$ yr$^{-1}$ delivered from the outer solar nebula. Since 100-km planetesimals are large enough not to have their orbits modified by gas drag but are too small to excite density waves at the Lindblad resonances and experience Type I migration (Ward 1997), the delivery mechanism is unclear and may require a nearby companion planet, which has been ruled out for HD 149026b. Finally, it is clear from the planet's non-negligible gas fraction that it spent some time in a gas-rich environment. If surrounded by gas while forming at $R < 0.05$ AU, HD 149026b would have migrated directly into the star in <1 yr. Due to the difficulty of constructing a physically realistic *in situ* formation scenario, we favor formation on a long-period orbit (~10 AU) followed by migration.

## 5. CONCLUSIONS

We have successfully reconstructed the growth history of the extrasolar planet HD 149026b based on its structural similarity with Saturn and the measured iron abundance of the host star. Not only have we brought a planet previously thought to require an exceptional formation mechanism under the purview of the core accretion theory, but we have demonstrated a method that can be applied to any exoplanet with a well-defined core mass orbiting a Solar-type star in order to discover its formation path:

1. Scale the solar nebula solid surface density up or down based on the star's heavy element abundance.

2. Constrain $R_0$ by examining similar planets with known formation histories and comparing the observed core mass to the isolation mass.

3. Using a core accretion simulation, check that (a) the planet forms in ~3 Myr or less and (b) the theoretical core/atmosphere mass ratio matches the observational values. If necessary, make small adjustments to $R_0$ and repeat.

Note that reliable measurements of core mass are critical to Step 2, as large core mass uncertainties allow $R_0$ to vary substantially. Even in the case of the ultra-dense HD 149026b, which unambiguously requires a large core, plausible core masses range between 45-110 $M_\oplus$ due to uncertainties in high-pressure equations of state, the distribution of heavy elements in the planet, atmospheric boundary conditions, and the assumed density of solid material. Furthermore, our core accretion model has uncertainties due to the aforementioned factors plus the range of possible formation distances and the core accretion rate. Our final core mass of 66 $M_\oplus$ is uncertain by at least 20 $M_\oplus$. However, the minimum core mass is still large enough to definitively exclude feeding zones inside 6 AU. We obtain a core/atmosphere mass ratio in the middle of the possible values and in excellent agreement with the 67 $M_\oplus$ quoted in the discovery paper (Sato et al. 2005) by invoking the hypothesis that HD 149026b's structural similarity with Saturn implies a similar formation mechanism.

Other candidate planets for a study of this type are OGLE-TR-111b (Pont et al. 2004) and OGLE-TR-113b (Bouchy et al. 2004), which have <30% core mass error bars (Burrows et al. 2007). Furthermore, the CoRoT and Kepler transit searches will jointly discover >100 planets with masses in the Neptune-Saturn range (19-95 $M_\oplus$) (Aigrain et al. 2008; Beatty & Gaudi 2008), which have high solid/gas ratios and therefore low core-mass uncertainties. High-precision transit photometry combined with a straightforward method for reconstructing the planet growth process will finally enable astronomers to derive robust formation histories for a statistically significant extrasolar planet population.

Support for S.D.R's work was provided by NASA through the Spitzer Space Telescope Fellowship Program. P.B. received support from the NASA Origins Grant NNX08AH82G. The authors thank Chas Beichman for helpful comments.

**FIGURES**

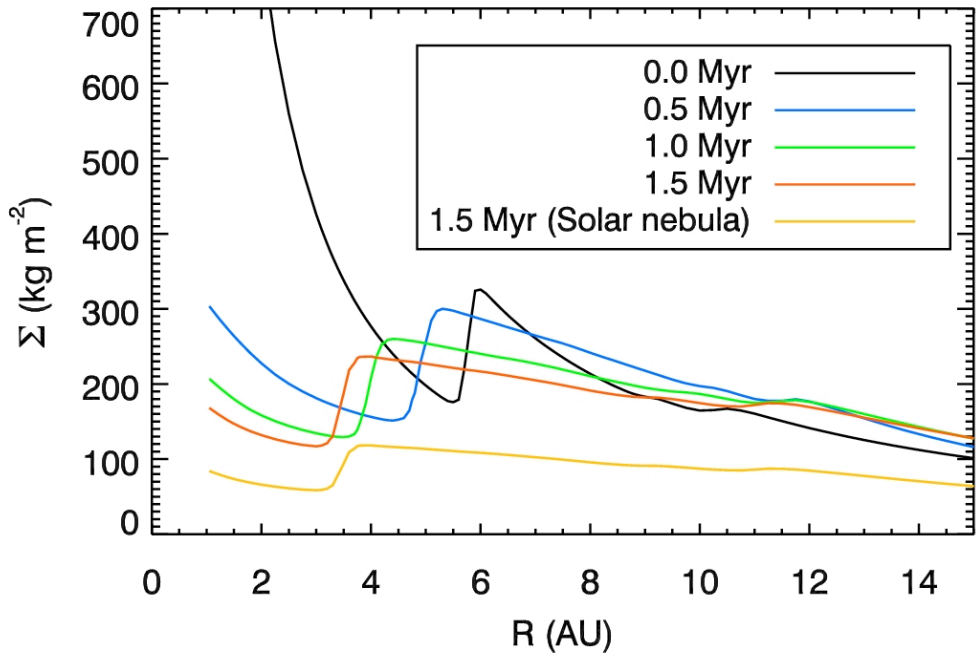

**Figure 1:** Solid surface density in the HD 149026 protostellar disk. Solid surface density curves Σ(R) (solid surface density as a function of radius) are plotted at four different times during the earliest evolutionary stage of the disk, 0-1.5 Myr (million years), when planetesimals form and agglomerate into planets. For reference, we also include the solar nebula surface density profile at t=1.5 Myr. The sharp spike in Σ(R) that begins at 6 AU and moves inward as the disk cools is the ice line.

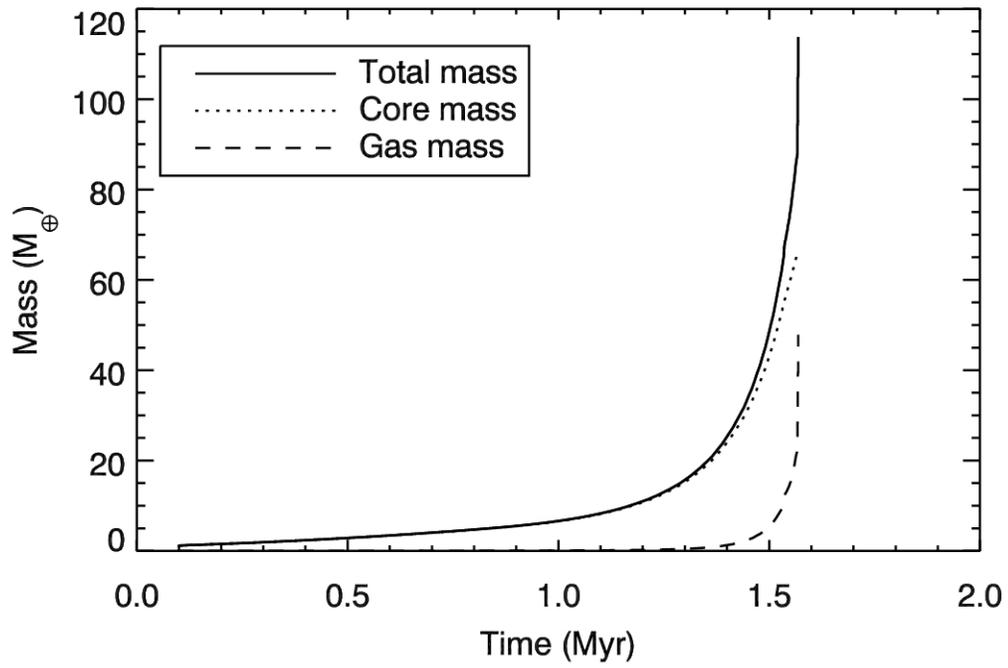

**Figure 2:** The growth of HD 149026b. At t = 1.57 Myr, when the planet reaches its current mass of 115 $M_\oplus$ (Earth masses), the core mass is 66 $M_\oplus$, in excellent agreement with the observed transit depth.

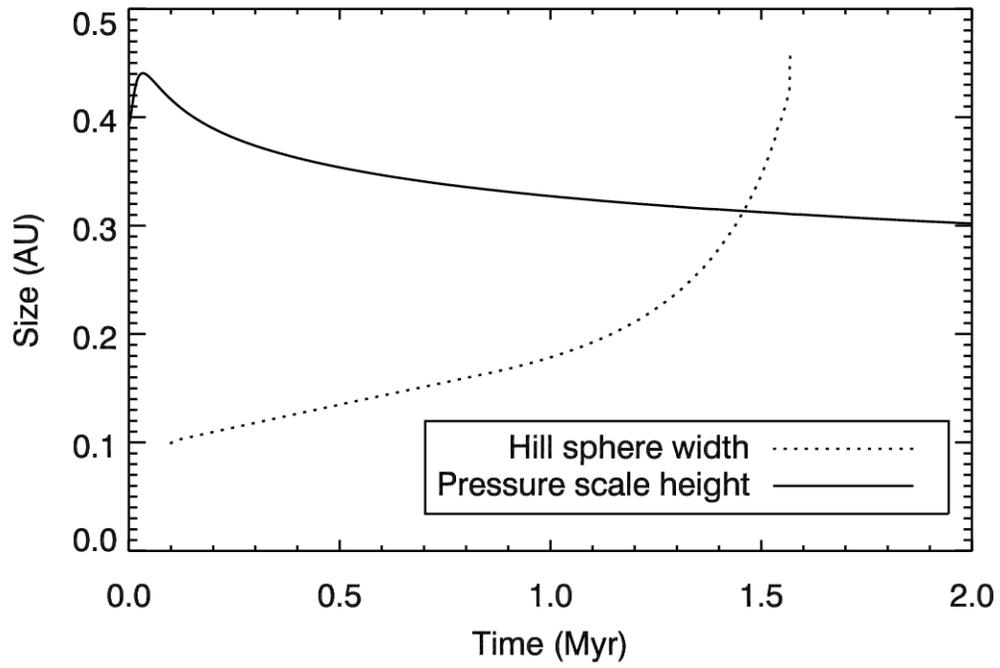

**Figure 3:** Hill sphere width and disk pressure scale height as a function of time. A gap in the disk gas opens when the two quantities become equal at t = 1.45 Myr, signaling the beginning of the Type II migration epoch.